
%
%
\input phyzzx
\tolerance=1000
\sequentialequations
\def\rl{\rightline}

\def\r#1{$\bf#1$}

\def\t1{{\tilde 1}}

\def\AEF{A.E. Faraggi}
\def\DVN{D. V. Nanopoulos}

\def\NPB#1#2#3{Nucl. Phys. B{\bf#1} (19#2) #3}
\def\PLB#1#2#3{Phys. Lett. B{\bf#1} (19#2) #3}
\def\PRD#1#2#3{Phys. Rev. D{\bf#1} (19#2) #3}
\def\PRL#1#2#3{Phys. Rev.Lett.{\bf#1} (19#2) #3}
\def\PRT#1#2#3{Phys. Rep. {\bf#1} (19#2) #3}

\def\IJMP#1#2#3{Int. J. Mod. Phys. A{\bf#1} (19#2) #3}

\def\l{\langle}
\def\r{\rangle}

\REF\mup{H. Georgi and I. N. McArthur, Harvard University Report HUTP--81/A001
(1981), unpublished; D. Kaplan and A. Manohar, \PRL{56}{86}{2004}.}
\REF\KIM{J.E. Kim, \PRT{150}{87}{1}; H.Y. Cheng, \PRT{158}{88}{1}.}
\REF\FNY{\AEF, D.V. Nanopoulos and K. Yuan, \NPB{335}{90}{347}.}
\REF\TOP{\AEF, \PLB{274}{92}{47}.}
\REF\EU{\AEF, \PLB{278}{92}{131} .}
\REF\SLM{\AEF, \PRD{47}{93}{5021}; \NPB{387}{92}{239}, hep-th/9208024.}
\REF\FFF{I. Antoniadis, C. Bachas, and C. Kounnas,
\NPB{289}{87}{87}; I. Antoniadis and C. Bachas,
\NPB{298}{88}{586}; H. Kawai, D.C. Lewellen, and S.H.-H. Tye,
\PRL{57}{86}{1832}; \PRD{34}{86}{3794};
\NPB{288}{87}{1}; R. Bluhm, L. Dolan, and P. Goddard,
\NPB{309}{88}{33}.}
\REF\FM{A.E. Faraggi, \NPB{407}{93}{57}, hep-ph/9301220.}
\REF\FOC{\AEF, IASSNS--77/93, hep-ph/9311312.}
\REF\KLN{S. Kalara, J. Lopez and D.V. Nanopoulos,
\PLB{245}{91}{421}; \NPB{353}{91}{650}.}
\REF\NRT{\AEF, \NPB{403}{93}{101}, hep-th/9208023.}
\REF\DSW{M. Dine, N. Seiberg and E. Witten,
\NPB{289}{87}{585};
J.J. Atick, L.J. Dixon and A. Sen, \NPB{292}{87}{109};
S. Cecotti, S. Ferrara and M. Villasante, \IJMP{2}{87}{1839}.}
\REF\CKM{A.E. Faraggi and E. Halyo, \PLB{307}{93}{305}, hep-ph/9301261;
         WIS--93/35/APR--PH, Nucl. Phys. {\bf B}, in press, hep-ph/9306235.}
\REF\ART{J. Lopez and \DVN, \NPB{338}{90}{72};
         J. Rizos and K. Tamvakis, \PLB{251}{90}{369};
         I. Antoniadis, J. Rizos and K. Tamvakis, \PLB{278}{92}{257}.}
\REF\GCU{\AEF, \PLB{302}{93}{202}, hep-ph/9301268.}
\REF\HALYO{E. Halyo, WIS--93/98/SEP--PH, hep-ph/9311300.}

\singlespace
\rl{IASSNS--HEP--93/82}
\rl{\today}
\rl{T}
\normalspace
\smallskip
\titlestyle{\bf{Light Fermion Masses in Superstring Derived Standard--like
Models}}
\author{Alon E. Faraggi
{\footnote*{e--mail address: faraggi@sns.ias.edu}}}
\smallskip
\centerline {School of Natural Sciences}
\centerline {Institute for Advanced Study}
\centerline {Olden Lane}
\centerline {Princeton, NJ 08540}

\titlestyle{ABSTRACT}

I discuss the suppression of the lightest generation fermion mass terms
in realistic superstring standard--like models in the free fermionic
formulation. The suppression of the mass terms is a consequence of horizontal
symmetries that arise due to the $Z_2\times Z_2$ orbifold
compactification. In a specific model I investigate the possibility of
resolving the strong CP puzzle by a highly suppressed up quark mass.
In some scenarios the up quark mass may be as small as
$10^{-8}MeV$. I show that in the specific model the suppression of the
up quark mass is incompatible with the requirement of a nonvanishing
electron mass. I discuss how this situation may be remedied.

\singlespace
\vskip 0.5cm
\nopagenumbers
\pageno=0
\endpage
\normalspace
\pagenumbers

\bigskip
\centerline{\bf Introduction}

Electroweak precision data from LEP indicate that the top quark may be found
in the mass range $110-200GeV$.
On the other hand the up and down quark masses
are of the order of $O(1MeV)$, while the well known electron mass is
$0.5MeV$. This vast separation of scales is one of the clues to the physics
beyond the standard model. In a theory of electroweak symmetry breaking the
expected mass of the top quark is rather natural as arising from  a
renormalizable operator with a Yukawa coupling, $\lambda_t$, of order one.
On the other hand the mass of the lightest generation states require Yukawa
couplings of the order $10^{-5}\lambda_t$. It may also be that
$\lambda_u\approx0$ is consistent with current algebra results and that the
mass of the up quark arises from nonperturbative strong interaction effects
rather than the value of the high energy parameter, thus providing a solution
to the strong CP problem [\mup]. In this paper I discuss the problem of the
suppression of the lightest generation mass terms in the context of realistic
superstring derived standard--like models. I discuss the possible
solution to the strong CP problem by the suppression of the up quark Yukawa
coupling in the context of these models.
To suppress CP violation in strong interactions requires
$\theta_{\hbox{tot}}(z/(1+z))<10^{-9}$ where
$\theta_{\hbox{tot}}=\theta_{\hbox{QCD}}+\theta_{\hbox{quarks}}$
and $z=m_u/m_d$ [\KIM]. I argue that in
some scenarios the up quark mass can be as small as $10^{-8}MeV$.

The superstring standard--like models [\FNY,\TOP,\EU,\SLM]
are constructed in the free fermionic formulation [\FFF].
To study the suppression of the lightest generation mass
terms I focus on the model that was presented in Ref. [\EU].
The standard--like models are generated
by sets of eight basis vectors,
$\{{\bf 1},S,b_1,b_2,b_3,\alpha,\beta,\gamma\}$. The set
$\{{\bf 1},S,b_1,b_2,b_3,2\gamma\}$ is common to all the realistic
models in the free fermionic formulation. The set
$\{{\bf 1},S,{\bf1}+b_1+b_2+b_3,2\gamma\}$ generates a toroidal compactified
model with $N=4$ space--time supersymmetry and
$SO(12)\times SO(16)\times SO(16)$ gauge symmetry. The vectors $b_1$ and
$b_2$ correspond to moding out the six dimensional torus by a $Z_2\times Z_2$
discrete symmetry with standard embedding, [\FM,\FOC].
The vectors $\alpha$, $\beta$, $\gamma$ differ
between models and correspond to different choices of Wilson line in the
orbifold language. The various choices of vectors $\alpha$, $\beta$, $\gamma$
and of the phases $c\left(\matrix{\alpha,\beta,\gamma\cr
                                   {\bf1},S,b_j\cr}\right)$
fix the physical spectrum and determine the low energy effective theory
of the superstring standard--like models.

The full massless spectrum together with the quantum numbers were
given in Ref. [\EU]. Here I summarize briefly the states that play a role in
the fermion mass matrices.
The gauge group after all GSO projections have been applied is
$\{SU(3)\times SU(2)\times U(1)_{B-L}\times U(1)_{T_{3_R}}\times U(1)^6\}_{_O}
\times\{SU(5)_H\times SU(3)_H\times U(1)^2\}_{_H}$,
where the first curly brackets
correspond to the observable gauge group that arises from the first $SO(16)$
times $SO(12)$. The second curly brackets arises from the second $SO(16)$.
The sectors $b_1$, $b_2$ and $b_3$ correspond to
the three twisted sectors of the orbifold model and produce three
16 of $SO(10)$ decomposed under
$SU(3)\times SU(2)\times U(1)_{B-L}\times U(1)_{T_{3_R}}$ with charges
under the horizontal symmetries.

The Neveu--Schwarz (NS) sector corresponds to the untwisted sector and
produces in addition to the gravity and gauge multiplets three pairs of
electroweak
scalar doublets $\{h_1, h_2, h_3, {\bar h}_1, {\bar h}_2, {\bar h}_3\}$,
three pairs of $SO(10)$ singlets with observable $U(1)$ charges,
$\{\Phi_{12},\Phi_{23},\Phi_{13},{\bar\Phi}_{12},
{\bar\Phi}_{23}, {\bar\Phi}_{13}\}$, and three scalars that are singlets
of the entire four dimensional gauge group, $\xi_1,\xi_2,\xi_3$.

The sector ${S+b_1+b_2+\alpha+\beta}$ ($\alpha\beta$ sector)
produces in addition to one pair of
electroweak doublets, $h_{45},{\bar h}_{45}$, and one pair of color
triplets, seven pairs of $SO(10)$
singlets with horizontal $U(1)$ charges, $\{\Phi_{45},{\bar\Phi}_{45},
\Phi_{1,2,3}^\pm,{\bar\Phi}_{1,2,3}^\pm\}$.

In addition to the states from these sectors, which transform solely under
the observable gauge group, the neutral states from the sectors $b_j+2\gamma$
and the sectors $b_{1,2}+b_3+\alpha+\gamma$ play a role in the fermion mass
matrices. The sectors $b_j+2\gamma$ produce $16$ vector representation
of the hidden $SO(16)$ gauge group decomposed under
$SU(5)_H\times SU(3)_H\times U(1)^2$,
$\{T_{1,2,3},{\bar T}_{1,2,3},V_{1,2,3},{\bar V}_{1,2,3}\}$.
These states are singlets of the observable $SO(10)$ gauge group but are
charged under the horizontal $U(1)^6$ charges. The states from the
sectors  $b_{1,2}+b_3+\alpha+\gamma$,
$\{H_{13},H_{14},H_{17},H_{18},H_{19},H_{20},H_{23},H_{24},H_{25},H_{26}\}$,
are standard model singlets but carry $U(1)_{Z^\prime}$ charge,
where  $U(1)_{Z^\prime}$ is the $U(1)$ inside $SO(10)$ that is orthogonal
to the electroweak hypercharge.

The cubic level superpotential and higher order nonrenormalizable terms
in the superpotential are obtained by calculating correlators between
vertex operators,
$A_N\sim\langle V_1^fV_2^fV_3^b\cdot\cdot\cdot V_N^b\rangle,$
where $V_i^f$ $(V_i^b)$ are the fermionic (scalar)
components of the vertex operators.
The nonvanishing terms must be invariant under all the symmetries of the
string models and must satisfy all the string selections rules [\KLN].
To obtain the correct ghost charge $(N_b-1)$ of the bosonic vertex
operators have to be picture changed from the $-1$ ghost picture to the
$0$ ghost picture. The invariance under the global left--moving $U(1)$
symmetries and the Ising model correlators must be checked after all
picture changing operations have been performed. The invariant terms
are extracted by using a simple FORTRAN code. In Ref. [\NRT] I discussed
the properties of the standard--like models which simplify the analysis
of nonrenormalizable terms.

The cubic level superpotential is given by,
$$\eqalignno{W&=\{(
{u_{L_1}^c}Q_1{\bar h}_1+{N_{L_1}^c}L_1{\bar h}_1+
{u_{L_2}^c}Q_2{\bar h}_2+{N_{L_2}^c}L_2{\bar h}_2+
{u_{L_3}^c}Q_3{\bar h}_3+{N_{L_3}^c}L_3{\bar h}_3)
+{{h_1}{\bar h}_2{\bar\Phi}_{12}}
+{h_1}{\bar h}_3{\bar\Phi}_{13}\cr
&\qquad
+{h_2}{\bar h}_3{\bar\Phi}_{23}
+{\bar h}_1{h_2}{\Phi_{12}}
+{\bar h}_1{h_3}{\Phi_{13}}
+{\bar h}_2{h_3}{\Phi_{23}}
+\Phi_{23}{\bar\Phi}_{13}{\Phi}_{12}
+{\bar\Phi}_{23}{\Phi}_{13}{\bar\Phi}_{12}
+{\bar\Phi}_{12}({\bar\Phi}_1^+{\bar\Phi}_1^-\cr
&\qquad
+{\bar\Phi}_2^+{\bar\Phi}_2^-
+{\bar\Phi}_3^+{\bar\Phi}_3^-)
+{\Phi_{12}}(\Phi_1^-\Phi_1^+
+\Phi_2^-\Phi_2^+
+\Phi_3^-\Phi_3^+)+{1\over2}\xi_3(\Phi_{45}{\bar\Phi}_{45}
+h_{45}{\bar h}_{45}+D_{45}{\bar D}_{45}\cr
&\qquad+\Phi_1^+{\bar\Phi}_1^++
\Phi_1^-{\bar\Phi}_1^-+\Phi_2^+{\bar\Phi}_2^++\Phi_2^-{\bar\Phi}_2^-
+\Phi_3^+{\bar\Phi}_3^++\Phi_3^-{\bar\Phi}_3^-)
+h_3{\bar h}_{45}\Phi_{45}+{\bar h}_3h_{45}{\bar\Phi}_{45}\}\cr
&\qquad
+\{{1\over2}[\xi_1(H_{19}H_{20}+
H_{21}H_{22}+H_{23}H_{24}+H_{25}H_{26})
+\xi_2(H_{13}H_{14}+H_{15}H_{16}+H_{17}H_{18})]\cr
&\qquad
+{\bar\Phi}_{23}H_{24}H_{25}
+{\Phi}_{23}H_{23}H_{26}+h_2H_{16}H_{17}
+{\bar h}_2H_{15}H_{18}
+{e_{L_1}^c}{H_{10}}{H_{27}}
+{e_{L_2}^c}{H_8}{H_{29}}
+({V_1}{H_9}\cr
&\qquad
+{V_2}{H_{11}}){H_{27}}+{V_6}{H_5}{H_{29}}
+{{\bar\Phi}_{45}}{H_{17}}{H_{24}}
+{D_{45}}{H_{18}}{H_{21}}
+{h_{45}}{H_{16}}{H_{25}}\}\quad&(1)\cr}$$
where a common normalization constant ${\sqrt 2}g$ is assumed.
{}From Eq. (2) it is seen that only $+{2\over3}$ charged quarks
obtain a cubic level mass term. This result arises due to the assignment
of boundary conditions in the vector $\gamma$ [\SLM].
Mass terms for $-{1\over3}$ and for
charged leptons must be obtained from nonrenormalizable terms.
The light Higgs spectrum is determined by the massless eigenstates of the
doublet Higgs mass matrix. The doublet mass matrix consists of the terms
$h_i{\bar h}_j\langle\Phi^n\rangle$, and is defined by
$h_i(M_h)_{ij}{\bar h}_j$, $i,j=1,2,3,4$
where $h_i=(h_1,h_2,h_3,h_{45})$
and ${\bar h}_i=({\bar h}_1,{\bar h}_2,{\bar h}_3,{\bar h}_{45})$.
At the cubic level of the
superpotential the Higgs doublets mass matrix is given by,
$$M_h={\left(\matrix{0&{\bar\Phi}_{12}&{{\bar\Phi}_{13}}&0\cr
                     \Phi_{12}&0&{{\bar\Phi}_{23}}&0\cr
                     {\Phi_{13}}&\Phi_{23}&0&{\Phi_{45}}\cr
                      0&0&{\bar\Phi}_{45}&0\cr}\right).}\eqno(2)$$
The superstring standard--like models contain an ``anomalous'' $U(1)$
gauge symmetry. The ``anomalous'' $U(1)$ symmetry generates a Fayet--Iliopolous
D--term at the one loop level that breaks supersymmetry at the Planck scale
and destabilize the vacuum [\DSW]. Supersymmetry is restored by giving a VEV
to some standard model singlets in the spectrum along F and D flat directions.
In the standard--like models, it has been found
that we must impose [\EU,\TOP,\NRT],
$$\langle{\Phi_{12},{\bar\Phi}_{12},\xi_3}\rangle=0,\eqno(3)$$
and that $\Phi_{45}$, and ${\bar\Phi}_{13}$ or ${\bar\Phi}_{23}$,
must be different from zero. From this result it follows that in any
flat F and D solution, $h_3$ and ${\bar h}_3$ obtain a Planck scale mass.
This result is a consequence of the symmetry of the vectors $\alpha$ and
$\beta$ with respect to the $b_1$ and $b_2$ sectors [\NRT].
The implication is that $h_3$ and ${\bar h}_3$ do not contribute to
the light Higgs representations. Consequently, the mass terms for the
states from the sector $b_3$ will be suppressed.

At the cubic level of the superpotential there are two pairs of light Higgs
doublets which may consist of combinations of $\{h_1,h_2,h_{45}\}$
and $\{{\bar h}_1,{\bar h}_2,{\bar h}_{45}\}$.
At the nonrenormalizable level of the superpotential, additional
non vanishing entries in the Higgs mass matrix appear [\NRT,\FM],
rendering one additional pair supermassive. The light Higgs representations
typically consist of ${\bar h}_1$ or ${\bar h}_2$ and $h_{45}$,
depending on the additional nonvanishing terms in the Higgs mass
matrix [\NRT]. In the analysis of nonrenormalizable terms I search
for any terms that include $(h_1,h_2,h_{45})$ and
$({\bar h}_1,{\bar h}_2,{\bar h}_{45})$ and thus do not make an assumption
as to what are the specific light Higgs combinations. The suppression
of the light fermion masses will be shown to be independent
of this choice. However, to examine whether one can obtain models
in which the strong CP problem is resolved by a sufficiently suppressed
up quark mass,
I will make the assumption that the light Higgs
representations are ${\bar h}_1$ and $h_{45}$.

Among the realistic models in the free fermionic formulation,
the standard--like
models have the unique property that there are three and only three
chiral generations. Therefore, the identification of the three light
generations is unambiguous.

\bigskip
\centerline{\bf 2.  Light fermion mass terms}

In Ref. [\FM] it was shown that the global left--moving horizontal symmetry
$U(1)_{\ell_3}$ forbid the formation of terms of the form
$f_3f_3h\phi^n$
or $f_3f_3{\bar h}\phi^{n}$, where $f_3$ are fermions from the
sectors $b_3$, $h$ and ${\bar h}$ are combinations of
$\{h_1,h_2,h_{45}\}$ and $\{{\bar h}_1,{\bar h}_2,{\bar h}_{45}\}$
respectively, and $\phi^{n}$ is a combination of $SO(10)$ singlet
fields from the Neveu--Schwarz sector and the sector $b_1+b_2+\alpha+\beta$.
In this paper I extend the analysis to the case where $\phi^{n}$
include scalar fields from the sectors $b_j+2\gamma$ and
$b_{1,2}+b_3+\alpha+\beta$.

At the quintic level the following mass terms are obtained
$$\eqalignno{&d_2Q_2h_{45}{\bar\Phi}_2^-\xi_1,{\hskip .2cm}
       e_2L_2h_{45}{\bar\Phi}_2^+\xi_1,{\hskip .2cm}
       d_1Q_1h_{45}{\Phi}_1^+\xi_2,{\hskip .2cm}
       e_1L_1h_{45}{\Phi}_1^-\xi_2&(4a)\cr
      &u_2Q_2({\bar h}_{45}\Phi_{45}{\bar\Phi}_{23}+
  {\bar h}_1{\bar\Phi}_i^+{\bar\Phi}_i^-),{\hskip .2cm}
      u_1Q_1({\bar h}_{45}\Phi_{45}{\bar\Phi}_{13}+
  {\bar h}_2{\Phi}_i^+{\Phi}_i^-)&(4b)\cr
      &(u_2Q_2h_2+u_1Q_1h_1)
  {{\partial W}\over{\partial\xi}_3}.&(4c)\cr}$$
At order $N=6$ we obtain mixing terms for $-{1\over3}$ charged quarks,
$$\eqalignno{
&d_3Q_2h_{45}\Phi_{45}V_3{\bar V_2},{\hskip .2cm}
d_2Q_3h_{45}\Phi_{45}V_2{\bar V_3},{\hskip .2cm}
d_3Q_1h_{45}\Phi_{45}V_3{\bar V_1},{\hskip .2cm}
d_1Q_3h_{45}\Phi_{45}V_1{\bar V_3},&(5)\cr}$$
and for charged leptons
$$\eqalignno{
&e_3L_2h_{45}\Phi_{45}T_3{\bar T_2},{\hskip .2cm}
e_2L_3h_{45}\Phi_{45}T_2{\bar T_3},{\hskip .2cm}
e_3L_1h_{45}\Phi_{45}T_3{\bar T_1},{\hskip .2cm}
e_1L_3h_{45}\Phi_{45}T_1{\bar T_3},&(6)\cr}$$
At order $N=7$ we obtain in the down quark sector,
$$\eqalignno{
&d_2Q_1h_{45}\Phi_{45}(V_1{\bar V_2}+V_2{\bar V_1})\xi_i,{\hskip .8cm}
d_1Q_2h_{45}\Phi_{45}(V_1{\bar V_2}+V_2{\bar V_1})\xi_i,&(7a,b)\cr
&d_1Q_3h_{45}\Phi_{45}V_3{\bar V_1}\xi_2,{\hskip 2.5cm}
d_3Q_1h_{45}\Phi_{45}V_1{\bar V_3}\xi_2,&(7c,d)\cr
&d_2Q_3h_{45}\Phi_{45}V_3{\bar V_2}\xi_1,{\hskip 2.5cm}
d_3Q_2h_{45}\Phi_{45}V_2{\bar V_3}\xi_1,&(7e,f) \cr}$$
where $\xi_i=\{\xi_1,\xi_2\}$.
In the up quark sector we obtain,
$$\eqalignno{
&u_1Q_2{\bar h}_1\Phi_{45}\{{\bar\Phi}_2^-(T_1{\bar T_2}+T_2{\bar T_1})+
                          {\bar\Phi}_1^+(V_1{\bar V_2}+V_2{\bar V_1)}\}&(8a)\cr
&u_2Q_1{\bar h}_1\Phi_{45}\{{\bar\Phi}_1^-(T_1{\bar T_2}+T_2{\bar T_1})+
                          {\bar\Phi}_2^+(V_1{\bar V_2}+V_2{\bar V_1})\}&(8b)\cr
&u_1Q_2{\bar h}_2\Phi_{45}\{{\Phi}_2^+(T_1{\bar T_2}+T_2{\bar T_1})+
                            {\Phi}_1^-(V_1{\bar V_2}+V_2{\bar V_1})\}&(8c)\cr
&u_2Q_1{\bar h}_2\Phi_{45}\{{\Phi}_1^+(T_1{\bar T_2}+T_2{\bar T_1})+
                            {\Phi}_2^-(V_1{\bar V_2}+V_2{\bar V_1})\}&(8d)\cr
&u_3Q_1{\bar h}_1\Phi_{45}\{{\bar\Phi}_1^-T_1{\bar T_3}+
                            {\bar\Phi}_3^+V_3{\bar V_1}\}{\hskip .5cm}
 u_1Q_3{\bar h}_1\Phi_{45}\{{\bar\Phi}_3^-T_1{\bar T_3}+
                            {\bar\Phi}_1^+V_3{\bar V_1})\}&(8e)\cr
&u_3Q_1{\bar h}_2\Phi_{45}\{{\Phi}_3^+T_1{\bar T_3}+
                            {\Phi}_1^-V_3{\bar V_1}\}{\hskip .5cm}
 u_1Q_3{\bar h}_2\Phi_{45}\{{\Phi}_3^+T_1{\bar T_3}+
                            {\Phi}_1^-V_3{\bar V_1})\}&(8f)\cr
&u_3Q_2{\bar h}_1\Phi_{45}\{{\bar\Phi}_2^-T_2{\bar T_3}+
                            {\bar\Phi}_3^+V_3{\bar V_2}\}{\hskip .5cm}
 u_2Q_3{\bar h}_1\Phi_{45}\{{\bar\Phi}_3^-T_2{\bar T_3}+
                            {\bar\Phi}_2^+V_3{\bar V_2})\}&(8g)\cr
&u_3Q_2{\bar h}_2\Phi_{45}\{{\Phi}_2^+T_2{\bar T_3}+
                            {\Phi}_3^-V_3{\bar V_2})\}{\hskip .5cm}
 u_2Q_3{\bar h}_2\Phi_{45}\{{\Phi}_3^-T_2{\bar T_3}+
                            {\Phi}_2^-V_3{\bar V_2})\}&(8h)\cr}$$

{}From the terms in Eqs. (4--8) we can construct fermion mass matrices that
lead
to quark mass and mixing spectrum of the correct order of magnitude [\CKM].
However, diagonal mass terms for the states from the sector $b_3$
do not appear in the equations above. Potential diagonal mass terms
for the lightest generation states are of the form $Q_3d_3h\phi^{n}$
and $Q_3u_3{\bar h}\phi^{n}$, where $h$ and ${\bar h}$ are combinations of
$\{h_1,h_2,h_{45}\}$ and $\{{\bar h}_1,{\bar h}_2,{\bar h}_{45}\}$
respectively, and $\phi^{n}$ is a string of standard model singlets.
The standard model singlets divide into several classes: (\romannumeral1)
$SO(10)$ singlets from the Neveu--Schwarz sector and the sector
$S+b_1+b_2+\alpha+\beta$. (\romannumeral2) $SO(10)$ singlets from the sectors
$b_j+2\gamma$. (\romannumeral3) States that carry $U(1)_{Z^\prime}$ charges
from the sectors $b_{1,2}+b_3+\alpha+\beta$.

Invariance under the left--moving global $U(1)_{\ell_3}$ symmetry forbids
the formation of terms $f_3f_3h{\phi}^{n}$ and
$f_3f_3{\bar h}\phi^{n}$, where $\phi^{n}$ are
restricted to class ({\romannumeral1}) singlets.
The argument goes as follows. The fermions from the sector
$b_3$ carry $U(1)_{\ell_3}={1\over2}$. The bosons from the Neveu--Schwarz
sector $\{h_3,{\bar h}_3,\Phi_{12},{\bar\Phi}_{12},\xi_3\}$ carry
$U(1)_{\ell_3}=-1$, and all the other states from the NS and $\alpha\beta$
sectors have $U(1)_{\ell_3}=0$. The charges of the twisted and untwisted
states under the symmetries $U(1)_{\ell_{1,2,3}}$ are due to the underlying
$Z_2\times Z_2$ orbifold compactification.
Since $h_3$ and ${\bar h}_3$
are supermassive, to form a potential mass term that is invariant
under $U(1)_{\ell_3}$, we must tag to $f_3f_3$, a Higgs state that
is neutral under $U(1)_{\ell_3}$ and one or more of
$\{\Phi_{12},{\bar\Phi}_{12},\xi_3\}$. However,
the $U(1)_{\ell_3}$ charges of $\{\Phi_{12},{\bar\Phi}_{12},\xi_3\}$
are changed to zero by picture changing and
therefore we cannot form a term that is invariant under
$U(1)_{\ell_3}$ with only class ({\romannumeral1}) singlets.
Thus, we have to examine terms that include class ({\romannumeral2}) and
({\romannumeral3}) singlets.
Below I focus on a scenario with ${\bar h}_1$ as the light Higgs that
couples to $+{2\over3}$ charged quarks.
A search up to order $N=9$ shows that terms
that include only class ({\romannumeral1}) and ({\romannumeral2})
singlets do not appear up to order $N=9$. At order $N=9$
we obtain for example in the up quark sector,
$$\eqalignno{
&Q_3u_3{\bar h_1}\Phi_{45}\Phi_{45}\{V_3{\bar V_2}{\bar T_3}T_2+
				     V_2{\bar V_3}{\bar T_2}T_3\}&(9a)\cr
&Q_3u_3{\bar h_1}\Phi_{45}\xi_1\{
({\bar\Phi}_2^-\Phi_{23}+{\Phi}_2^+\Phi_{13}){\bar T}_2T_2+
({\bar\Phi}_2^+\Phi_{23}+{\Phi}_2^-\Phi_{13}){\bar V}_2V_2\}&(9f)\cr}$$
with additional terms of the form of Eq.(9) with ${\bar h}_1$ replaced
by ${\bar h}_2$ and ${\bar h}_{45}$ to make a total of 35 terms.

If we include terms that break $U(1)_{Z^\prime}$ then we obtain at the
quintic order,
$$Q_3u_3({\bar h}_2H_{24}H_{25}+{\bar h}_{45}H_{17}H_{24})\eqno(10)$$
and at order $N=6$
$$Q_3d_3\{({h}_1\Phi_3^++{h}_2{\bar\Phi}_3^-)H_{17}H_{24}+
          {h}_{45}{\bar\Phi}_3^-H_{24}H_{25}\}\eqno(11)$$
at order $N=7$ we obtain in the up quark sector,
$$\eqalignno{
&Q_3u_3\{{\bar h}_1{\bar\Phi}_i^+{\bar\Phi}_i^-+
{\bar h}_2(\Phi_i^\pm\Phi_i^\pm+\xi_2\xi_2+
\Phi_{13}{\bar\Phi}_{13})\}H_{24}H_{25}&(12a)\cr
&Q_3u_3{\bar h}_2\Phi_{45}\xi_2H_{18}H_{25}&(12b)\cr
&Q_3u_3{\bar h}_{45}\Phi_{45}(
{{\partial W_3}\over{\partial\xi_1}}+
{{\partial W_3}\over{\partial\xi_2}})\xi_i\xi_j&(12c)\cr}$$
Plus additional terms of the form $Q_3u_3({\bar h}_2+{\bar h}_{45})H^4$
where $\langle H\rangle$ breaks $U(1)_{Z^\prime}$.

The texture of the quark mass matrices is determined by the choice of singlet
VEVs. The singlet VEVs are constrained by the F and D flat constraints. In a
general solution we may obtain quark masses of order MeV. For example
assuming assuming $\l H\r\approx0$ and taking the F and D flat
direction from Ref.
[\CKM] the quark mass matrices take the form
$$M_u\sim\left(\matrix{&\epsilon
&{{V_3{\bar V}_2\Phi_{45}\bar \Phi_3^+}\over{M^4}} &0\cr
&{{V_3{\bar V}_2\Phi_{45}\bar \Phi_2^+}\over{M^4}}
&{{{\bar\Phi}_i^-\bar \Phi_i^+}\over{M^2}}
&{V_1{\bar V}_2\Phi_{45}\bar \Phi_2^+}\over{M^4} \cr
&0 &{V_1{\bar V}_2\Phi_{45}{\bar\Phi}_1^+}\over{M^4}
&1\cr}\right)v_1,\eqno(13)$$
and
$$M_d\sim\left(\matrix{&\epsilon
&{{V_3{\bar V}_2\Phi_{45}}\over{M^3}} &0\cr
&{{V_3{\bar V}_2\Phi_{45}\xi_1}\over{M^4}}
&{{{\bar\Phi}_2^-\xi_1}\over{M^2}} &{V_1{\bar V}_2\Phi_{45}\xi_i}\over{M^4} \cr
&0 &{V_1{\bar V}_2\Phi_{45}\xi_i}\over{M^4}
&{{\Phi_1^+\xi_2}\over{M^2}}\cr}\right)v_2,\eqno(14)$$
where $v_1=\l \bar h_1 \r$,  $v_2=\l{h}_{45}\r$ and $\epsilon\sim0$.
The up, down quark masses and the Cabibbo angle are given by
$$m_u\approx{{(M_u)_{12}(M_u)_{21}}\over{(M_u)_{22}}}=
{\l{V_3{\bar V}_2\Phi_{45}\r^2\l{\bar\Phi}^+_3{\bar\Phi}_2^+\r}\over
{M^6{\l{\bar\Phi}_i^+{\bar\Phi}_i^-}\r}}v_1\eqno(15)$$ and
$$\eqalignno{
&m_d \approx {{(M_d)_{12}(M_d)_{21}}\over{(M_d)_{22}}}=
{\l{V_3{\bar V}_2\Phi_{45}\r^2\l\xi_1\r}\over
{M^6{\l{\bar\Phi}_2^-\r\l\xi_1}\r}}v_2\qquad\qquad&(16a)\cr
&\sin\theta_c\approx{{(M_d)_{12}}\over(M_d)_{22}}=
{{\l V_3{\bar V}_2\Phi_{45}\r}\over{M\l\bar\Phi}_2^-\r\l\xi_1\r}&(16b)\cr}$$
To make an attempt at a numerical estimate of the up and down quark masses
I take the F and D flat solution that was found in Ref. [\CKM],
$$\eqalignno{
&{1\over3}\vert{{\l{\bar V}_2\r}\over{M}}\vert^2=
 {1\over2}\vert{{\l{V_3}\r}\over{M}}\vert^2=
 {1\over6}\vert{{\l\Phi_{45}\r}\over{M}}\vert^2=
          \vert{{\l{\bar\Phi}_3^+\r}\over{M}}\vert^2=
         2\vert{{\l{\bar\Phi}_2^+\r}\over{M}}\vert^2=
{g^4\over{16\pi^2}}&(17)\cr}$$
and ${{\l{\bar\Phi}_i^+{\bar\Phi}_i^-\r}\over{M^2}}\approx0.01$.
With this solution and with $\l\xi_1\r\sim{1\over4}M$ we obtain Cabibbo
angle and down quark mass of the correct order of magnitude [\CKM].
{}From Eq. (15) and (17), we observe that
Taking $g\sim0.8$ the unification scale [\GCU], and $v_1\sim100GeV$
we obtain naively $m_u\sim0.01MeV$ from this solution. In a general
solution we may assume $\l\phi\r/M\sim0.1$, which yields $m_u\sim0.1MeV$.

The interesting observation regarding the quark mass matrices is that
for particular choices of flat directions the textures of the down and up
mass matrices is different [\CKM].
This entails the possibility that the diagonal entry $(M_u)_{11}$ one of the
non--diagonal entries in the up quark mass matrix $(M_u)_{12}$
or $(M_u)_{21}$ vanish up to some order [\CKM]. Consequently, it is possible
that the mass
of the lightest up quark state vanishes or is highly suppressed, while the
down quark and the Cabibbo angle are of the correct order of magnitude.
This is an interesting possibility as it may provide a solution to the
strong CP problem.

First, I estimate the possible contribution from the order $N=9$ terms
Eqs. (9). I impose that only one state from a given sector $b_j+2\gamma$
gets a VEV, $\{V_1,{\bar V}_2,V_3\}$. Therefore,
the only terms that contribute are those that contain the
condensates of the hidden $SU(5)$ gauge group.
The bilinear hidden sector condensates produce a suppression factor
that is given by
$$({\Lambda_5\over{M}})^2=\exp({{2\pi}\over{b}}{{(1-\alpha_0)}\over
\alpha_0})\eqno(18)$$
where $b={1\over2}n_5-15$. For $n_5=6$ and $\alpha_0=(1/20-1/25)$,
$\Lambda_5\sim(10^{12}-10^{14})GeV$.
Taking, $\Lambda_5\sim 10^{14}GeV$ and
${{\phi}/{M}}\approx({1/{10}}-{1/{25}})$
I estimate the contribution
of the $N=9$ order term to be in the range
$m_u\sim(10^{-6}-10^{-8})MeV$, where I took $v_1=\l{\bar h}_1\r\sim100GeV$
and multiplied by a factor of ten to account for the multiplicity of terms.
If we take $\Lambda_5\sim10^{13}GeV$ then we obtain
$m_u\sim(10^{-9}-10^{-11})MeV$.

If in addition we assume that $\Lambda_{Z^\prime}$ is suppressed,
say $\Lambda_{Z^\prime}\leq10^{10}GeV$, then the contribution from
the terms that break $U(1)_{Z^\prime}$ is suppressed by at least
$({\Lambda_{Z^\prime}/{M}})^2\sim10^{-16}$. Thus, these terms are
sufficiently suppressed and produce $m_u\sim10^{-11}MeV$. In Ref. [\NRT]
it was argued that VEVs that break $U(1)_{Z^\prime}$ have to be
suppressed because of the constraint that higher order
nonrenormalizable terms should not spoil the cubic level F and D flat solution.
The constraints on the $U(1)_{Z^\prime}$ breaking VEVs are investigated
further in Ref. [\HALYO], where it is suggested that already at the cubic
level the F flatness constraints restrict the $U(1)_{Z^\prime}$
breaking VEVs. However, lacking an understanding of the SUSY breaking
mechanism, the $U(1)_{Z^\prime}$ VEVs may still be large, say of the
order $10^{14}GeV$, and produce soft SUSY breaking terms that are in accord
with the naturalness constraints on the SUSY spectrum.
In this case the up quark mass
will be of the order $10^{-3}MeV$, not small enough to resolve the strong
CP puzzle. Thus, to obtain $m_u$ sufficiently small we have to assume
$\Lambda_{Z^\prime}\le10^{11}GeV$.

In Eq. (9) we obtained nonvanishing terms of the form
$u_3Q_3{\bar h}_1\phi^n$ where $\phi$ are class ({\romannumeral1})
and ({\romannumeral2}) singlets only. These terms are suppressed
by $(\Lambda_5/M)^2\sim(10^{-8}-10^{-10})$. However, there may exist
higher order terms $(N>9)$ with only class ({\romannumeral1})
and ({\romannumeral2}) singlets that are not suppressed by the hidden
sector condensation scale. In this case the singlets $\phi$ belong to the set
$\l\phi\r\in\{\{NS\},\{\alpha\beta\},V_1,{\bar V}_2,V_3\}$. At order
$N=11$ we obtain the terms,
$$Q_3u_3{\bar h}_1(\Phi_{45}V_3{\bar V}_2)^2(\Phi_2^-{\bar\Phi}_3^++
\Phi_3^-{\bar\Phi}_2^+).\eqno(19)$$
Inserting the numerical values from the solution in Eq. (17) and with
$(\l\Phi_3^-\r/M)=(3g^2)/(\sqrt{10}4\pi)$ [\CKM], we obtain
$\lambda_u\sim10^{-9}$ or $m_u\approx10^{-4}MeV$. Thus, to construct
model in which $m_u$ is sufficiently suppressed we have to impose
$\l\Phi_3^-\r\approx0$. Imposing $\l\Phi_3^-\r=0$ and
$\l{\bar\Phi}_3^+\r=0$ then guarantees that the diagonal mass terms
are suppressed up to order $N=11$. At order $N=12$ there are no
invariant terms of the desired form. At order $N=13$ we obtain
$$Q_3u_3{\bar h}_1(\Phi_{45}V_1{\bar V}_2)^2(
\Phi_{23}{\bar\Phi}_2^+{\bar\Phi}_1^++\Phi_{13}{\bar\Phi}_1^+\Phi_2^-
+\Phi_{13}{\bar\Phi}_2^+\Phi_1^-)(\xi_1+\xi_2)\eqno(20)$$
Thus, in the best case scenario the diagonal mass terms will be suppressed
up to order $N=13$. Inserting the numerical values from the solution in
Ref. [\CKM] and with $g\sim(0.8-0.7)$ we obtain
$\lambda_u\sim(10^{-12}-12^{-13})$ or $m_u\sim(10^{-7}-10^{-8})MeV$.

Next I examine the contribution from the nondiagonal terms.
If we choose a flat F and D direction solution with
$\l{\bar\Phi}_3^+\r=0$ then $(M_u)_{12}$ in Eq. (13)
vanishes up to order $N=7$,
while from Eq. (4a) we observe that to give mass to the $\mu$ lepton we must
have $\l{\bar\Phi}_2^+\r\ne0$.
Thus, to examine whether it is possible to obtain
$m_u$ sufficiently suppressed to resolve the strong CP problem I focus on the
terms that contribute to $(M_u)_{12}$. Eq. (8g) shows that at order
$N=7$ there is contribution from condensates of the hidden $SU(5)$ gauge
group. This contribution is estimated to be $(M_u)_{12}<10^{-10}v_1$,
where I have taken $\Lambda_5\sim10^{14}GeV$ and $\l\phi\r\sim{1/10}$.
{}From Eq. (15) this contributes less than $O(10^{-7}MeV)$ to $m_u$.
Next, I examine terms that contribute to $(M_u)_{12}$ which are not suppressed
by hidden sector condensates.
These terms must be of the form
$u_3Q_2{\bar h}_1V_3{\bar V}_2\phi^n$, where $n=1,\cdots,N-5$, and
$\l\phi\r\in\{\{NS\},\{\alpha\beta\},V_1,{\bar V}_2,V_3\}$.
At order $N=8$ there are no potential terms contributing to
$u_3Q_2{\bar h}_1V_3{\bar V}_2$.
At order $N=9$ we obtain the following potential terms
$$\eqalignno{
&u_3Q_2{\bar h}_1\Phi_{45}\Phi_{45}V_3{\bar V}_2{\bar\Phi}_3^+
(\xi_i\xi_i+\xi_1\xi_2+\Phi_{13}{\bar\Phi}_{13}+\Phi_{23}{\bar\Phi}_{23}+
{{\partial W_3}\over{\partial\xi_3}})&(21a)\cr
&u_3Q_2{\bar h}_1\Phi_{45}\Phi_{45}V_3{\bar V}_2\Phi_3^-
({{\partial W_3}\over{\partial\Phi_{12}}})&(21b)\cr}$$
Plus additional terms that are suppressed by
$({{\Lambda_{Z^\prime}}/{M}})^2$.
The requirement ${\bar\Phi}_3^+=0$ imposes that the terms in Eq.(21a)
vanish identically. The terms in Eq. (21b) are suppressed by the cubic level
F flatness constraint
$({{\partial W_3}/{\partial\Phi_{12}}})=0.$
Thus, the order $N=9$ terms are suppressed by at least
$10^{-4}({{M_{\hbox{SUSY}}}/{M_{Pl}}})^2\leq10^{-12}$.
At order $N=10$ there are no terms that contribute to
$(M_u)_{12}$ which are not suppressed by hidden sector condensates.
At order $N=11$ and higher there will be many additional terms.
Invariance under $U(1)_{r_6}$ necessitates that either ${\bar\Phi}_3^+$ or
$\Phi_3^-$ appear in the correlators $u_3Q_2{\bar h}_1V_3{\bar V}_2\phi^n$.
Thus, all these terms vanish if we impose $\l{\bar\Phi}_3^+\r\approx0$ and
$\l\Phi_3^-\r\approx0$ on the F and D flat solution.

In summary, the diagonal mass terms for the lightest generation states
are suppressed due to the horizontal symmetry $U(1)_{\ell_3}$. The suppression
of the lightest generation states results from the basic structure of the
vectors $\alpha$ and $\beta$ with respect to the sectors $b_1$, $b_2$, $b_3$
and the resulting constraints on flat directions [\FM].
Therefore, the suppression
of the lightest generation states is expected to be a general characteristic
of the standard--like models. In fact, in the flipped $SU(5)$ superstring
model, where vectors similar to $\alpha$ and $\beta$ are constructed, similar
constraints on flat directions are obtained [\ART].
The up and down quark mass matrices possess a different texture. This enables
obtaining Cabibbo angle and down quark mass of the correct order of magnitude
while suppressing the up quark mass. With the assumption that ${\bar h}_1$
and $h_{45}$ are the light Higgs representations, to obtain an up quark mass
that is highly suppressed the following constraints must be imposed:
(1) Only one state from each sector $b_j+2\gamma$ can obtain a VEV.
(2) VEVs which break $U(1)_{Z^\prime}$ have to be suppressed with
$\Lambda_{Z^\prime}<10^{11}GeV$. (3) $SU(5)$ condensation scale of order
$O(10^{13}GeV)$. (4) $\l{\bar\Phi}_3^+\r\approx0$ and $\l\Phi_3^-\r\approx0$.
In this case the leading contribution to $(M_u)_{12}$ is from terms
that are suppressed by hidden sector condensates and the contribution to the
diagonal mass term is from order $N=13$ terms which are not
suppressed by hidden sector condensates. In this best case scenario
the up quark mass can be as small as $10^{-8}MeV$.

Next I address the problem of the electron mass. The electron mass is
obtained from diagonal terms, $e_3L_3h_{45}\l\phi\r^n$, and nondiagonal terms,
$e_{2,3}L_{3,2}h_{45}\l\phi\r^n$. Eq. (6) shows that order $N=6$
there are terms that are suppressed by the hidden sector condensation scale
$(\Lambda_5/M)^2$. Assuming a solution with $(M_e)_{22}\sim 10^{-3}$ and
taking $\l h_{45}\r\approx100GeV$ produces $m_e\sim10^{-8}MeV$. Clearly
too small. At order $N=6$ an additional term that breaks
$U(1)_{Z^\prime}$ is obtained,
$$e_3L_3h_{45}{\bar\Phi}_3^+H_{24}H_{25}.\eqno(22)$$
Assuming, $\Lambda_{Z^\prime}\sim10^{14}GeV$, $\l\phi\r\sim1/10$ and
$\l h_{45}\r\sim100GeV$ we get $m_e\sim10^{-4}MeV$. At order $N=8$ we get
$$e_2L_3h_{45}\Phi_{45}\Phi_2^-{\bar\Phi}_3^+V_3{\bar V}_2\qquad\qquad
  e_3L_2h_{45}\Phi_{45}\Phi_3^-{\bar\Phi}_2^+V_3{\bar V}_2\eqno(23a,b)$$
Taking $\l\phi\r\sim M/10$ yields $m_e\sim10^{-2}MeV$. Up to order
$N=12$ all diagonal terms, $e_3L_3$,
are suppressed by at least $(\Lambda_5/M)^2$
or $(\Lambda_{Z^\prime}/M)^2$. At order $N=12$ we obtain terms that are
suppressed only by singlet VEVs of order $M/10$. However,
invariance under $U(1)_{r_6}$ dictates that to all order of
nonrenormalizable terms $\Phi_3^-$ and ${\bar\Phi}_3^+$ must appear
in the correlators of the form $e_3L_3h_{45}\phi^n$, where
$\phi\in\{\{NS\},\{\alpha\beta\},V_1,{\bar V}_2,V_3\}$.

{}From the discussion above it is evident that in this model the suppression of
$m_u$ and an electron mass of order $(0.1MeV)$ are incompatible
as the first requires $\l{\bar\Phi}_3^+\r\approx0$ and
$\l\Phi_3^-\r\approx0$, while the second requires the opposite.
Replacing ${\bar h}_1$ by ${\bar h}_2$ will produce similar
results as it does not affect the invariance of the
correlators under $U(1)_{r_6}$.
A remedy to this situation may be obtained by modifying the phases
$c\left(\matrix{       b_j\cr
                       \alpha,\beta,\gamma\cr}\right)$. The
left-- right--moving horizontal symmetries $U(1)_{{\ell,r}_{4,5,6}}$
fix the invariant nonrenormalizable terms. The horizontal symmetries
and the Ising model operators arise from the internal fermionic states
$\{y,\omega\vert{\bar y}{\bar\omega}\}$ which correspond to the six dimensional
compactified space in an orbifold formulation [\FM,\FOC]. Modifying the
phases $c\left(\matrix{b_j\cr \alpha,\beta,\gamma\cr}\right)$  modifies
the GSO projections and consequently the charges under these horizontal
symmetries. Thus for example it may be possible to choose phases that will
produce electron nondiagonal mass terms at order $N=7$
that are suppressed only by singlet VEVs of order $M/10$, while
the corresponding up quark mass terms will be pushed to higher
orders. A similar dependence of nonrenormalizable terms on the boundary
conditions of the fermionic states $\{y,\omega\vert{\bar y},{\bar\omega}\}$
was found in the case of quartic and quintic order bottom quark
mass terms [\TOP,\SLM]. Thus, this is a viable possibility and merits further
investigation.

\vfill\eject
\centerline{\bf Acknowledgments}
I thank Edi Halyo for useful discussions.
This work is supported by an SSC fellowship.
\refout

\vfill
\eject

\end